\documentclass[a4paper,11pt]{article}
\pdfoutput=1 

\usepackage{jheppub} 

\usepackage[T1]{fontenc} 

\usepackage{epsfig}
\usepackage{graphicx}
\usepackage{amsmath}
\usepackage{amsfonts,amssymb}
\usepackage{color}

  \usepackage{slashed}
  \usepackage{amsthm}
\usepackage{float}
\usepackage{bm}
\usepackage{verbatim}

\usepackage{epstopdf} 

\renewcommand{\matrix}[4]{\left(\begin{array}{cc} #1 & #2 \\ #3 & #4\end{array}\right)}

\title{\boldmath $\rho$ condensation and physical parameters}

\author[a]{Marco Frasca,\note{Corresponding author.}}

\affiliation[a]{Via Erasmo Gattamelata, 3 \\
             00176 Roma (Italy)}

\emailAdd{marcofrasca@mclink.it}

\abstract{Recently we showed how a non-local Nambu-Jona-Lasinio model comes out from QCD in the low-energy limit. In this way, it is possible to fix all the free parameters of the model with physical ones. We use this approach to derive a local limit to the Nambu-Jona-Lasinio model with the parameters those obtained from QCD in order to fix the physical parameters of $\rho$ condensation. $\rho$ condensation is a consequence of the highly non-trivial behavior of the QCD vacuum in presence a very strong magnetic field giving rising to superconductive behavior in quark matter. Determination of the proper parameters for this state can be an important helpful guide to identify it experimentally.}

\begin{document} 
\maketitle
\flushbottom

\section{Introduction}

The study of interacting matter in strong magnetic field is a hot activity field in physics (see the review \cite{Kharzeev:2013jha}). These results have impact in a wide variety of areas. A great interest arose about this matter for the studies on QCD \cite{Ruggieri:2013cya,Gatto:2010pt,Gatto:2010qs,Fukushima:2010fe,Menezes:2008qt,Frasca:2011zn,Gatto:2012sp,Fukushima:2012vr,Allen:2013lda} also using extensive lattice computations \cite{Bali:2011qj,D'Elia:2011zu,Bali:2012zg,Bruckmann:2013oba,D'Elia:2010nq,Endrodi:2013cs}. The reason is that experiments with heavy ions at RHIC or LHC can easily give rise to such high magnetic fields applied to strong interactions. So, it is mandatory to get an understanding of the behavior of the vacuum of QCD in such situations. This is generally accomplished using phenomenological models that we are confident should describe rather accurately the behavior of QCD in this extreme states \cite{Klevansky:1989vi,Shushpanov:1997sf,Kabat:2002er,Inagaki:2003yi,Cohen:2007bt,Fraga:2008qn,Agasian:2008tb,Mizher:2010zb}.

Following this track, recently Chernodub put forward the proposal that, at a given critical magnetic field, $\rho$ vectors could form a condensate proper to a superconductive state \cite{Chernodub:2010qx,Chernodub:2011mc,Chernodub:2012tf}. This will characterize a quantum phase transition where it is not the varying of temperature to cause a transition but rather some other control physical parameter that in this case is the magnetic field \cite{Sondhi:1997zz}. This is an exciting possibility and there has been some studies on the lattice to try to evidence it \cite{Braguta:2011hq,Hidaka:2012mz}. In the paper by Hidaka and Yamamoto \cite{Hidaka:2012mz}, it was pointed out that a theorem by Vafa and Witten exists \cite{Vafa:1983tf} that forbids the existence of condensates
for $\rho$ mesons like this one, breaking the diagonal subgroup of the global isospin group $U(1)_{I_3}$ of QCD, that leads to the appearance of a massless Nambu-Goldstone boson in the spectrum.
%
They support this conclusion using lattice computations but, on the other side, leave an open door for models like the Nambu-Jona-Lasinio one. This contradicts the lattice computations provided in \cite{Braguta:2011hq} but, on the same track, a recent paper showed how the Vafa-Witten theorem does not seem to apply to this case \cite{Chernodub:2012zx,Chernodub:2013uja}. The question remains open about lattice evidence and further studies are needed to clarify this point. It is also interesting to see a possible experimental evidence for such a superconductive state in QCD.

The existence of a condensate for the $\rho$ vector bosons was inferred using phenomenological models. In a first case \cite{Chernodub:2010qx}, it was considered the model yielded in \cite{Djukanovic:2005ag} and then, a Nambu-Jona-Lasinio model was postulated \cite{Chernodub:2011mc}. Recently, a non-local Nambu-Jona-Lasinio has been to shown to be a consequence of QCD in the low-energy limit \cite{Frasca:2008zp,Kondo:2010ts,Frasca:2011bd,Frasca:2012eq,Frasca:2012iv}. This means that is possible to prove, starting from QCD, the existence of this quantum phase transition in QCD. This is so because all the parameters of the Nambu-Jona-Lasinio model are directly determined from QCD itself. Besides, this model has been largely discussed in literature and so, we are in a position to strongly corroborate the conclusions drawn in \cite{Chernodub:2011mc}.

The idea we apply in this paper is quite simple. The pole of the $\rho$ propagator is well-known from the Nambu-Jona-Lasinio model \cite{Ebert:1994mf,Bernard:1993rz}. This can be straightforwardly generalized to the case of finite temperature and magnetic field \cite{Menezes:2008qt}. Then, it can be shown that the mass of the $\rho$ develops a singularity at a critical magnetic field for zero temperature at increasing magnetic field. This is the hallmark of a quantum phase transition. Vector meson dominance also follows as a consequence and corrections dependent on the temperature can be easily computed but these are exponentially damped. The existence of this pole support the effect postulated by Chernodub in his papers. We expect a smaller critical field as, differently from the phenomenological model for vector mesons given in \cite{Djukanovic:2005ag}, here we have bounded states of quarks.

The paper is so structured. In Sec.\ref{sec1} we derive the infrared limit of QCD giving also comparison with lattice data for the gluon propagator and the form factor for an instanton liquid. In Sec.\ref{sec2} we describe $\rho$ condensation in the simplest phenomenological model to show how this effect firstly come about. In Sec.\ref{sec3}, we compute the value of the critical field using the result presented in \cite{Chernodub:2011mc}. In Sec.\ref{sec4}, we prove the existence of the quantum phase transition for the $\rho$ at increasing magnetic field finding the value of the critical field and also the corrections due to temperature as this lowers toward zero. Finally, in \ref{sec5} the conclusions are given.

\section{QCD in the infrared limit}
\label{sec1}

We briefly summarize the argument to derive a non-local Nambu-Jona-Lasinio for the sake of completeness. Details are presented elsewhere \cite{Frasca:2008zp,Frasca:2011bd,Frasca:2012eq,Frasca:2012iv}. Anyhow, with respect to the previous analysis we improve on the form of the propagator in agreement with the existence of exact solutions of the massless scalar field \cite{Frasca:2009bc}.

A prototypical field theory is a massless scalar field with a simple non-linearity given by
\begin{equation}
\label{eq:cphi}
   \Box\phi+\lambda\phi^3=j.
\end{equation}
The corresponding homogeneous equation admits an exact solution \cite{Frasca:2009bc} $\phi_0(x) = \mu\left(2/\lambda\right)^\frac{1}{4}{\rm sn}(p\cdot x+\theta,i)$ being {\rm sn} an elliptic Jacobi function and $\mu$ and $\theta$ two integration constants. In order for this solutions to hold, the following dispersion relation applies $p^2=\mu^2\sqrt{\lambda/2}$. We recognize here a free massive solution notwithstanding we started from a massless theory. We see that mass arises from the nonlinearities when $\lambda$ is taken to be finite rather than going to zero and so, standard perturbation theory just fails to recover it. Our aim is to solve the equation (\ref{eq:cphi}) in the limit $\lambda\rightarrow\infty$ and for our aims we consider an approach devised in the '80s \cite{Cahill:1985mh} taking $\phi$ as a functional of $j$ and expanding in powers of it. So, we put
\begin{equation}
\label{eq:ps}
   \phi[j]=\phi_0(x)+\int d^4x'\left.\frac{\delta \phi[j]}{\delta j(x')}\right|_{j=0}j(x')
   +\int d^4x'd^4x''\left.\frac{\delta^2 \phi[j]}{\delta j(x')\delta j(x'')}\right|_{j=0}j(x')j(x'')+\ldots
\end{equation}
being $\left.\frac{\delta \phi[j]}{\delta j(x')}\right|_{j=0}=\Delta(x-x')$ a solution to the nonlinear equation $\Box \Delta(x)+3\lambda[\phi_0(x)]^2\Delta(x)=\delta^4(x)$. So, this current expansion is meaningful as a strong coupling expansion. The interesting result here is the linear term of the Green function in the current expansion. When applied to a quantum field theory this will provide a Gaussian generating functional marking a trivial theory. So, we recognize that this set of classical solutions yields, at the leading order, a trivial theory. Such a theory is mathematically manageable notwithstanding the strong nonlinearity of the theory we started from. This program can be accomplished if we know how to get the Green function. This can be computed immediately yielding \cite{Frasca:2009bc}
\begin{equation}
\label{eq:green}
   \Delta(p)=\sum_{n=0}^\infty\frac{B_n}{p^2-m_n^2+i\epsilon}
\end{equation}
being
\begin{equation}
   B_n=(2n+1)^2\frac{\pi^3}{4K^3(i)}\frac{e^{-(n+\frac{1}{2})\pi}}{1+e^{-(2n+1)\pi}}
\end{equation}
and $m_n=(2n+1)(\pi/2K(i))\left(\lambda/2\right)^{\frac{1}{4}}\mu$ and $K(i)\approx 1.3111028777$ an elliptic integral, consistently with the idea of a strong coupling expansion. This holds provided one fixes the phase $\theta$ in the exact solution to $\theta_m=(4m+1)K(i)$. This identifies an infinite set of scalar field theories with a trivial infrared fixed point in quantum field theory.

We would like to apply all this machinery to Yang-Mills theory but in order to show this we need to have a set of classical solutions to work with also in this case. This set of solutions would grant a trivial infrared fixed point also for this theory. Such solutions exist. This can be seen starting from the equations of motion
\begin{equation}
\partial^\mu\partial_\mu A^a_\nu-\left(1-\frac{1}{\xi}\right)\partial_\nu(\partial^\mu A^a_\mu)+gf^{abc}A^{b\mu}(\partial_\mu A^c_\nu-\partial_\nu A^c_\mu)+gf^{abc}\partial^\mu(A^b_\mu A^c_\nu)+g^2f^{abc}f^{cde}A^{b\mu}A^d_\mu A^e_\nu = -j^a_\nu.
\end{equation}
and assuming again a current expansion. We note that the homogeneous equations can be solved by setting $A_\mu^a(x)=\eta_\mu^a\phi(x)$ being $\eta_\mu^a$ a set of constants. In this case we need to fix the gauge and we assume the Lorenz gauge being this equivalent to the Landau gauge in quantum field theory. So, the homogeneous equations collapse to $\partial^\mu\partial_\mu\phi+Ng^2\phi^3=-j_\phi$ and we have turned back to the previous scalar field theory (this is no more true for other gauges where the correspondence is just an asymptotic one \cite{Frasca:2009yp}). In this way, the gluon propagator in the Landau gauge is straightforwardly obtained from eq.(\ref{eq:green}) setting $\lambda=Ng^2$ and with a factor $\delta_{ab}\left(\eta_{\mu\nu}-p_\mu p_\nu/p^2\right)$. These solutions give us confidence that an analysis with a trivial infrared fixed point can also be performed in the case of a Yang-Mills theory. Indeed, that such a trivial infrared fixed point for the running coupling could exist is strongly supported by lattice computations as shown by the German group \cite{Bogolubsky:2009dc} from lattice at $64^4$ and $80^4$ with $\beta=5.7$ where the running coupling is seen to go to zero as momenta lower. A similar result was obtained by the French group with a different definition of the infrared running coupling \cite{Boucaud:2002fx}. This latter computation shows a perfect consistency with an instanton liquid model in agreement with the scenario we are depicting here.

\begin{figure}[H]
  \includegraphics{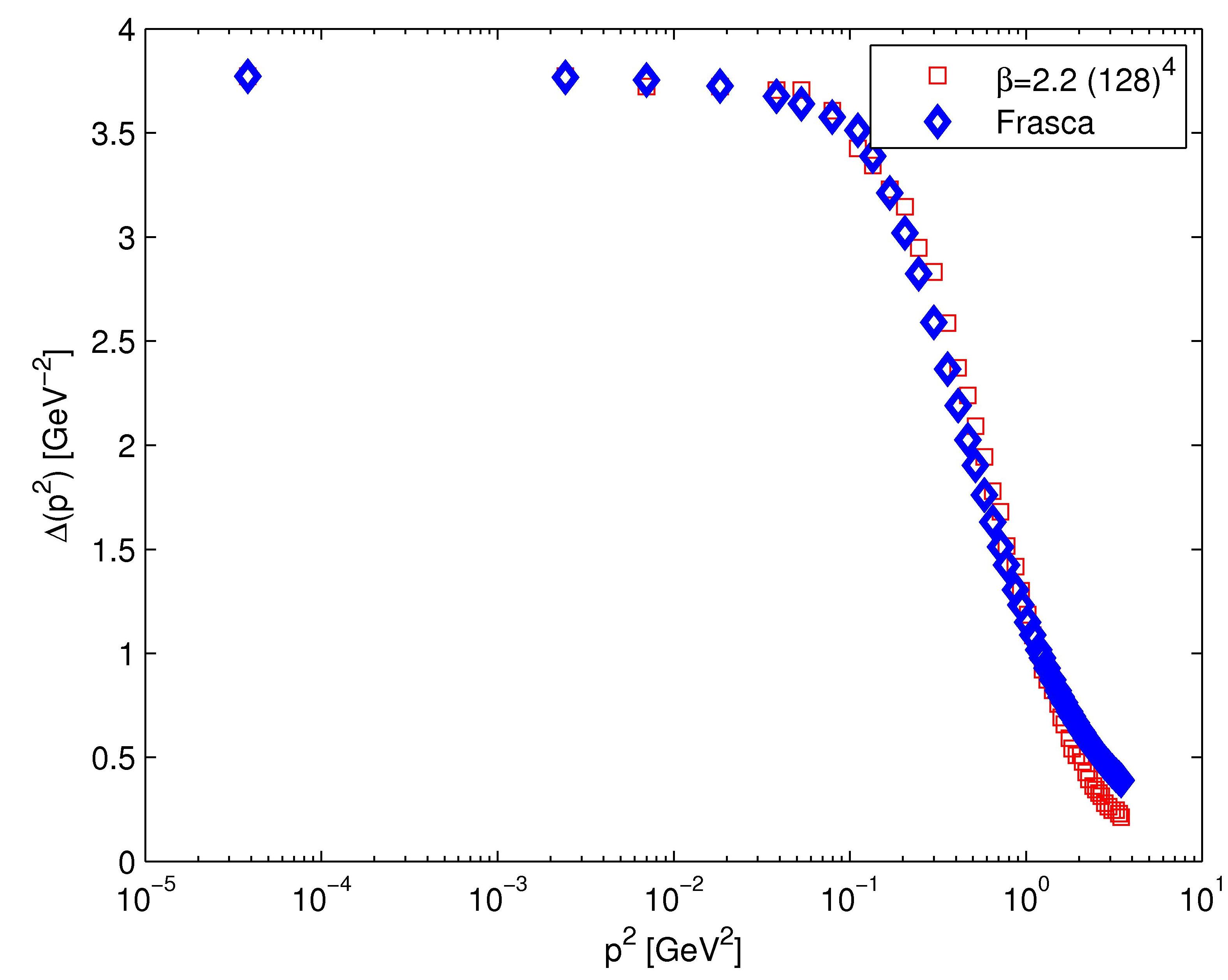}
  \caption{Comparison of our propagators with the lattice data for SU(2) given in \cite{Cucchieri:2007md} for $(128)^4$ points.\label{fig:cuccmend}}
\end{figure}

Moving to quantum field theory, the generating functional for the scalar field can be managed by rescaling the space-time coordinates as $x\rightarrow\sqrt{\lambda}x$ and with a strong coupling expansion $\phi=\sum_{n=0}^\infty\lambda^{-n}\phi_n$. Then, at the leading order we will have to solve the equation $\Box\phi_0+\lambda\phi_0^3=j$ that we now know how to manage. Then, the leading order is just a Gaussian generating functional with the propagator given by eq.(\ref{eq:green}) when use is made of the approximation in eq.(\ref{eq:ps}), next-to-leading order can be also computed. We arrive at the fundamental result that there exist an infinite set of massless scalar field theories in four dimensions that are infrared trivial \cite{Frasca:2009bc}. Mass spectrum is given by $m_n=(2n+1)(\pi/2K(i))\left(\lambda/2\right)^\frac{1}{4}\mu$ as expected, representing free particles with a superimposed spectrum of a harmonic oscillator. Now, turning the attention to Yang-Mills generating functional we realize that it takes the simple Gaussian form
\begin{equation}
     Z_0[j]=N\exp\left[\frac{i}{2}\int d^4x'd^4x''j^{a\mu}(x')D_{\mu\nu}^{ab}(x'-x'')j^{b\nu}(x'')\right].
\end{equation}
once we use the current expansion $A_\mu^a=\Lambda\int d^4x' D_{\mu\nu}^{ab}(x-x')j^{b\nu}(x')+O\left(1/\sqrt{N}g\right)+O(j^3)$ and the propagator in the Landau gauge $D_{\mu\nu}^{ab}(p)=\delta_{ab}\left(\eta_{\mu\nu}-\frac{p_\mu p_\nu}{p^2}\right)\Delta(p)$ being $\Delta(p)$ given by eq.(\ref{eq:green}). The spectrum in this case is that of free massive glue excitations with a superimposed spectrum of a harmonic oscillator. This is consistent with our initial observation of a trivial infrared fixed point for the running coupling. Similarly, considering the ghost field, applying our approximation on the gauge field through instanton solutions, this decouples at the leading order producing a free particle propagator for a massless field. These properties of the quantum Yang-Mills field describe the so-called ``decoupling solution'' \cite{Aguilar:2004sw,Boucaud:2006if,Frasca:2007uz} (see also \cite{Weber:2011nw} for a discussion). This solution is the one recovered in lattice computations \cite{Bogolubsky:2007ud,Cucchieri:2007md,Oliveira:2007px}. In order to see how our propagator is consistent with respect to lattice computations we show a comparison in Fig.\ref{fig:cuccmend}
The agreement is exceedingly good as expected.

\begin{figure}[H]
  \includegraphics{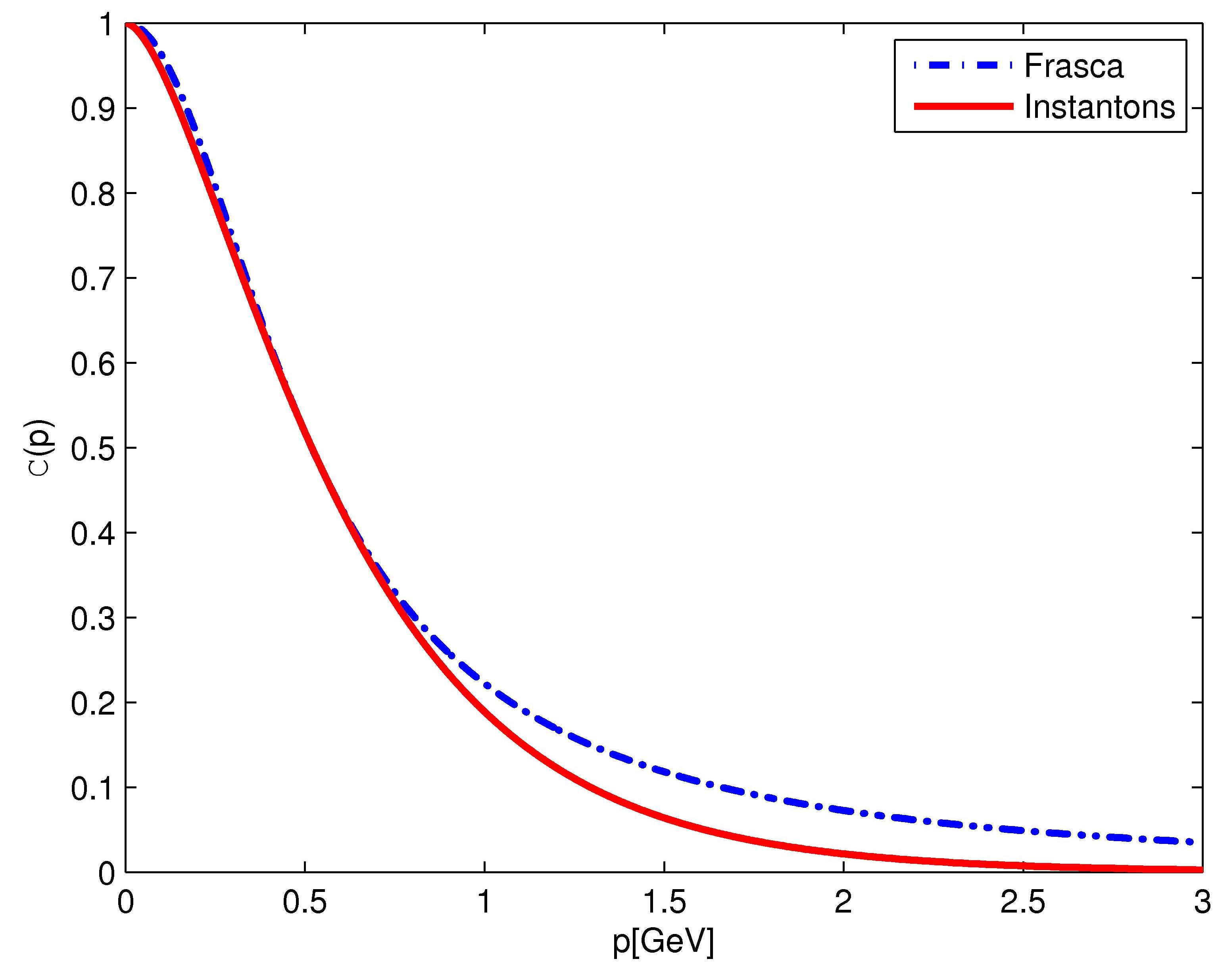}
  \caption{Comparison of our form factor with that provided in \cite{Schafer:1996wv} 
  for $\sqrt{\sigma}=0.417\ GeV$ and $d^{-1}=0.58\ GeV$.\label{fig:ff}}
\end{figure}

One can apply this low-energy behavior of Yang-Mills theory to the full QCD generating functional, with the propagator the one we just obtained, yielding for the action entering into it \cite{Frasca:2011bd,Frasca:2012eq,Frasca:2012iv}
\begin{equation}
      S=\int d^4x\left[\frac{1}{2}(\partial\sigma)^2-\frac{1}{2}m_0^2\sigma^2\right]+S_q
\end{equation}
where the $\sigma$ field arises from the gluon propagator in the Gaussian generating functional of the Yang-Mills action, neglecting higher order excited state in the superimposed harmonic oscillator spectrum, and gives the contribution from the mass gap of the theory, being $m_0=(\pi/2K(i))\sqrt{\tilde\sigma}$ and $\tilde\sigma$ is the string tension ($\approx (440\ MeV)^2$). For the quark fields one has
\begin{eqnarray}
\label{eq:njl}
      S_q&=&\sum_q\int d^4x\bar q(x)\left[i{\slashed\partial}-m_q-g\sqrt{\frac{B_0}{3(N_c^2-1)}}
      \eta_\mu^a\gamma^\mu\frac{\lambda^a}{2}\sigma(x)\right]q(x) \\  
     &-&g^2\int d^4x'\Delta(x-x')\sum_q\sum_{q'}\bar q(x)\frac{\lambda^a}{2}\gamma^\mu\bar q'(x')\frac{\lambda^a}{2}\gamma_\mu q'(x')q(x)
      +O\left(\frac{1}{\sqrt{N}g}\right)+O\left(j^3\right). \nonumber
\end{eqnarray}
Now, we are able to recover the non-local Nambu-Jona-Lasinio model given in \cite{Hell:2008cc} in the way yielded in \cite{Frasca:2011bd}, directly from QCD, provided the form factor is
\begin{equation}
\label{eq:Gp}
      {\cal G}(p)=-\frac{1}{2}g^2\Delta(p)=-\frac{1}{2}g^2\sum_{n=0}^\infty\frac{B_n}{p^2-(2n+1)^2(\pi/2K(i))^2\tilde\sigma+i\epsilon}
      =\frac{G}{2}{\cal C}(p)
\end{equation}
being $B_n$ obtained from eq.(\ref{eq:green}), ${\cal C}(0)=1$ and $2{\cal G}(0)=G$ the standard Nambu-Jona-Lasinio coupling, fixing in this way the value of $G$ through the gluon propagator. In Fig.\ref{fig:ff}, we compare this form factor both with the one from an instanton liquid \cite{Schafer:1996wv} that is
\begin{equation}
\mathcal{C}_I(p)=p^2\left\{\pi d^2 \dfrac{d}{d\xi}\big[I_0(\xi)K_0(\xi)-I_1(\xi)K_1(\xi)\big]\right\}^2\qquad\text{with } \xi=\frac{|p| d}{2} 
\end{equation}
being $I_n$ and $K_n$ Bessel functions. In the following we normalize this function to be 1 at zero momenta dividing it by ${\cal C}_I(0)$.
The result is strikingly good for the latter showing how consistently our technique represents Yang-Mills theory through instantons. We can safely conclude that the Nambu-Jona-Lasinio model is not confining due to the properties of its vacuum that is well represented by an instanton liquid. Confinement can be recovered adding higher order corrections to this model \cite{Frasca:2012iv}.

So, finally we write down the NJL action we will use in the following as was obtained from QCD
\begin{eqnarray}
\label{eq:njlok}
      S_{NJL}&=&\sum_q\int d^4x\bar q(x)\left[i{\slashed\partial}-m_q\right]q(x) \nonumber \\  
     &+&\int d^4x'{\cal G}(x-x')\sum_q\sum_{q'}\bar q(x)\frac{\lambda^a}{2}\gamma^\mu\bar q'(x')\frac{\lambda^a}{2}\gamma_\mu q'(x')q(x).
\end{eqnarray}
To put it into the standard form a Fierz transformation is needed. We will discuss this in sec.\ref{sec3} where also a constant magnetic field will be applied.

\section{$\rho$ condensation}
\label{sec2}

We give in the following a simple argument, as already presented in \cite{Chernodub:2010qx,Chernodub:2012tf}, that shows how in principle $\rho$ condensation can happen. An in depth analysis will be performed with the Nambu-Jona-Lasinio model in the next section.

The simplest way to see how a superconductive state, through $\rho$ condensation, could emerge in QCD with a strong magnetic field is to consider the Lagrangian \cite{Chernodub:2010qx}
\begin{eqnarray}
\label{eq:Lag} 
{\cal L} & = & -\frac{1}{4} \ F_{\mu\nu}F^{\mu\nu}
- \frac{1}{2} (D_{\mu}\rho_{\nu}-D_{\nu}\rho_{\mu})^\dagger (D^{\mu}\rho^{\nu}-D^{\nu}\rho^{\mu}) + m_\rho^2 \ \rho_\mu^\dagger \rho^{\mu} \nonumber
\\ & & 
 - \frac{1}{4} \rho^{(0)}_{\mu\nu} \rho^{(0) \mu\nu}{+}\frac{m_\rho^2}{2} \rho_\mu^{(0)}
\rho^{(0) \mu} +\frac{e}{2 g_s} F^{\mu\nu} \rho^{(0)}_{\mu\nu}\,, 
\end{eqnarray}
being $\rho_\mu = (\rho^{(1)}_\mu - i \rho^{(2)}_\mu)/\sqrt{2}$ and $\rho^{(0)}_\mu \equiv \rho^{(3)}_\mu$ the charged and neutral vector mesons made out of the components of the triplet of the $\rho$ field:
\begin{eqnarray}
\label{eq:rhofield}
\rho_{\mu} = 
\left(
\rho_{\mu}^{(1)},
\rho_{\mu}^{(2)},
\rho_{\mu}^{(3)} 
\right)^{T}\,.
\end{eqnarray}
The last term in Eq.~(\ref{eq:Lag}) describes a non-minimal coupling of the $\rho$ mesons to the electromagnetic field via the field strength $F_{\mu\nu} = \partial_{\mu}A_{\nu}-\partial_{\nu} A_{\mu}$ of the photon field $A_\mu$. The presence of the non-minimal coupling implies, in particular, the anomalously large value of the gyromagnetic ratio of the $\rho$ meson, $g = 2$. Both the covariant derivative $D_\mu = \partial_\mu + i g_s \rho^{(0)}_\mu - ie A_\mu$ and the strength tensor $\rho^{(0)}_{\mu\nu} = \partial_{\mu}\rho^{(0)}_{\nu}-\partial_{\nu}\rho^{(0)}_{\mu} - i g_s (\rho^\dagger_{\mu}\rho_{\nu}-\rho^\dagger_{\nu}\rho_{\mu})$ involve the $\rho\pi\pi$ coupling $g_s$ which has the known phenomenological value of $g_s \approx 5.88$.

In order to identify an order parameter and the appearance of a condensate, the strategy is to compute the energy density for the given Lagrangian with a constant magnetic field and to evaluate the potential when no dependence on space-time variable is present (homogeneous approximation). This is a common way to proceed, in a mean-field approximation, with the Landau model for second order phase transitions. Here we are coping with a quantum phase transition as the control parameter is not the temperature but rather the magnetic field \cite{Sondhi:1997zz}. We assume $F_{12}=B$ the constant magnetic field applied. It is not difficult to identify the energy density
\begin{eqnarray}
\label{eq:energy:density}
& & V\bigl(\rho_\mu,\rho^{(0)}_\nu\bigr) = \frac{1}{2} B^2 + \frac{g_s^2}{4}
\bigl[i \bigl(\rho_\mu^\dagger \rho_\nu - \rho_\nu^\dagger \rho_\mu\bigr)\bigr]^2\\
& &+ i e B \bigl(\rho_1^\dagger \rho_2 - \rho_2^\dagger \rho_1\bigr)
+ \frac{m_\rho^2}{2} {\bigl(\rho_\mu^{(0)}\bigr)}^2
+ m_\rho^2 \rho_\mu^\dagger \rho_\mu\,. \quad
\nonumber
\end{eqnarray}
where Euclidean sum on repeated indexes is implicit with indexes running from 1 to 3. The potential for $\rho_\mu^{(0)}$ is vanishing in 0 and this yields for the quadratic part of the Lagrangian
\begin{eqnarray}
\label{eq:rho:mu}
V^{(2)}(\rho_\mu,0) & = & i e B \bigl(\rho_1^\dagger \rho_2 - \rho_2^\dagger \rho_1\bigr) + m_\rho^2 \rho_\mu^\dagger \rho_\mu
\nonumber\\
& = & \sum_{a,b=1}^2\rho_a^\dagger {\cal M}_{ab} \rho_b + m_\rho^2 (\rho_0^\dagger \rho_0 + \rho_3^\dagger \rho_3)\,.
\end{eqnarray}
being the mass matrix
\begin{eqnarray}
\label{eq:cM}
{\cal M} =
\left(
\begin{array}{cc}
m_\rho^2 & i e B \\
- i e B & m_\rho^2
\end{array}
\right)\,.
\end{eqnarray}
This means that the physical states are identified after diagonalization of the mass matrix (\ref{eq:cM}) and these are
\begin{eqnarray}
\label{eq:rho:diag}
\mu_{\pm}^2 = m_\rho^2 \pm e B\,,
\qquad
\rho_{\pm} = \frac{1}{\sqrt{2}} (\rho_1 + i \rho_2)\,.
\end{eqnarray}
Turning back to the potential (\ref{eq:energy:density}), noticing that the minimum has $\rho_0=\rho_3=0$ and writing it in terms of the physical states, one gets
\begin{eqnarray}
\label{eq:V01}
V(\rho_+,\rho_-) = \frac{1}{2} B^2 & {+} & \frac{g^2_s}{2} \bigl(|\rho_+|^2 - |\rho_-|^2\bigr)^2\nonumber\\
&+&  \mu_+^2 |\rho_+|^2 + \mu_-^2 |\rho_-|^2\,.
\end{eqnarray}
This potential has the standard form for second order phase transitions. We choose for the ground state
\begin{eqnarray}
\label{eq:gs}
\rho_1 = - i \rho_2 = \rho\,, \qquad \rho_0 = \rho_3 = 0\,,
\end{eqnarray}
so that
\begin{eqnarray}
\label{eq:gs2}
V(\rho) = \frac{1}{2} B^2 + 2 (m_\rho^2 - e B) \, |\rho|^2 + 2 g_s^2 \, |\rho|^4\,.
\end{eqnarray}
This potential has exactly the form of that required for a second order phase transition but the parameter to be changed is the magnetic field. We are working at zero temperature and so, we are characterizing a quantum phase transition. We can immediately identify a critical magnetic field
\begin{equation}
   B_c=\frac{m_\rho^2}{e}
\end{equation}
that when it is overcome a condensate develops with a vacuum expectation value
\begin{equation}
   \langle\rho\rangle=\pm\sqrt{\frac{B-B_c}{2g_s^2}}.
\end{equation}
This confirms the possible existence of a phase transition with a $\rho$ condensate but, before we can be certain, a more in-depth analysis is needed using an effective model coming out from QCD: The Nambu-Jona-Lasinio model. The homogeneous approximation is a too strong one and this appears rather as a clue.

As said in the introduction, the apparent conflict with the Vafa-Witten theorem \cite{Vafa:1983tf} for the existence
of a $\rho$ meson 
condensate does not apply here \cite{Chernodub:2012zx,Chernodub:2013uja}.

\section{Computation of the critical magnetic field}
\label{sec3}

The aim of this section is to compute the critical magnetic field for the condensate to set in using the Nambu-Jona-Lasinio model obtained above and compare it with that yielded by Chernodub in \cite{Chernodub:2011mc}. 

So, let us consider the Lagrangian (\ref{eq:njlok}) for two flavors of quarks and applying an external magnetic field $B$
\begin{eqnarray}
\label{eq:njlA}
     {\cal L}_{NJL}&=&\frac{1}{2}(\partial\sigma)^2-\frac{1}{2}m_0^2\sigma^2\nonumber \\
     &+&\sum_{q=u,d}\bar q(x)\left[i{\slashed\partial} + e_q \, {\slashed {\cal A}}-m_q\right]q(x) \nonumber \\  
     &-&g^2\int d^4x'\Delta(x-x')\sum_{q=u,d}\sum_{q'=u,d}\bar q(x)\frac{\lambda^a}{2}\gamma^\mu q(x)
     \bar q'(x')\frac{\lambda^a}{2}\gamma_\mu q'(x').
\end{eqnarray}
being $e_u=+2e/3$ and $e_d=-e/3$ and ${\cal A}_\mu=\left(0,Bx_2/2,-Bx_1/2,0\right)$. The next step is to put this Lagrangian in a more standard form and this can be easily obtained with a Fierz rearrangement. This will preserve the original symmetries in the Lagrangian and we will consider
\begin{eqnarray}
\label{eq:njlF}
     {\cal L}_{NJL}&=&\frac{1}{2}(\partial\sigma)^2-\frac{1}{2}m_0^2\sigma^2\nonumber \\
     &+&\bar \psi(x)\left[i{\slashed\partial} + \hat e \, {\slashed {\cal A}}-\hat m\right]\psi(x) \nonumber \\  
     &+&\int d^4x'{\cal G}(x-x')
     \left[\bar\psi(x)\psi(x)\bar\psi(x')\psi(x')+\bar\psi(x)i\gamma^5{\bm\tau}\psi(x)\bar\psi(x')i\gamma^5{\bm\tau}\psi(x')\right. \nonumber \\
     &-&\left.\frac{1}{2}\bar\psi(x)\gamma^\mu{\bm\tau}_a\psi(x)\bar\psi(x')\gamma_\mu{\bm\tau}_a\psi(x')
     -\frac{1}{2}\bar\psi(x)\gamma^5\gamma^\mu{\bm\tau}_a\psi(x)\bar\psi(x')\gamma^5\gamma_\mu{\bm\tau}_a\psi(x')\right]
\end{eqnarray}
being $\psi = (u,d)^T$, $\hat e={\rm diag}(e_u,e_d)$, $\hat m={\rm diag}(m_u,m_d)$ and ${\bm\tau}_a=(I,{\bm\tau})$ Pauli matrices (we use subscript $a$ for vectors with four components). We can identify
\begin{equation}
    {\cal G}_S(x-x')={\cal G}(x-x') \qquad {\cal G}_V(x-x')=\frac{1}{2}{\cal G}(x-x').
\end{equation}
Four our aims, it is not convenient to work with a non-local model. We then are able to get the coupling constant of the Nambu-Jona-Lasinio model from eq.(\ref{eq:Gp}) to give ${\cal G}(0)=G=g^2/(2\tilde\sigma)$, completely defined in terms of QCD observables. The Lagrangian is now
\begin{eqnarray}
\label{eq:njlL}
     {\cal L}_{NJL}&=&\frac{1}{2}(\partial\sigma)^2-\frac{1}{2}m_0^2\sigma^2\nonumber \\
     &+&\bar \psi(x)\left[i{\slashed\partial} + \hat e \, {\slashed {\cal A}}-\hat m\right]\psi(x) \nonumber \\  
     &+&\frac{G}{2}
     \left[(\bar\psi(x)\psi(x))^2+(\bar\psi(x)i\gamma^5{\bm\tau}\psi(x))^2\right] \nonumber \\
     &-&\frac{G}{4}\left[(\bar\psi(x)\gamma^\mu{\bm\tau}_a\psi(x))^2
     +(\bar\psi(x)\gamma^5\gamma^\mu{\bm\tau}_a\psi(x))^2\right]
\end{eqnarray}
and can be bosonized in a standard way \cite{Ebert:1994mf,Bernard:1993rz} giving the effective field theory. One introduces the field $\sigma(x)=G\bar\psi(x)\psi(x)$ and ${\bm\pi}(x)=G\bar\psi(x)\gamma^5{\bm\tau}\psi(x)$ and
\begin{eqnarray}
\label{eq:matrix:U}
{\hat V}_\mu & \equiv & {\bm \tau}_a\cdot{\bm V}_\mu = \matrix{\omega_\mu + \rho^{0}_\mu}{\sqrt{2} \rho^+_\mu}{\sqrt{2} \rho^-_\mu}{\omega_\mu - \rho^{0}_\mu}\,,
\qquad
V_\mu^i=G\bar\psi \gamma_\mu \tau^i \psi\,,
\end{eqnarray}
[composed of the flavor-singlet coordinate-vector $\omega$--meson field $\omega_\mu$, and of the electrically neutral, $\rho^0_\mu \equiv \rho^3_\mu$, and charged, $\rho^\pm_\mu = (\rho^{1}_\mu \mp i \rho^{2}_\mu)/\sqrt{2}$, components of the $\rho$-meson triplet], and four pseudovector (axial) fields,
\begin{eqnarray}
\label{eq:matrix:A}
{\hat A}_\mu & \equiv & {\bm \tau}_a\cdot{\bm A}_\mu = \matrix{f_\mu + a^{0}_\mu}{\sqrt{2} a^+_\mu}{\sqrt{2} a^-_\mu}{f_\mu - a^{0}_\mu}\,, 
\qquad
A_\mu^i=G\bar\psi \gamma^5 \gamma_\mu \tau^i \psi\,.
\end{eqnarray}
This will yield
\begin{eqnarray}
\label{eq:LB}
   {\cal L}_B&=&\frac{1}{2}(\partial\sigma)^2-\frac{1}{2}m_0^2\sigma^2\nonumber \\
   &+&\bar\psi i{\cal D}\psi-\frac{1}{2G}\left({\sigma^2}+{\bm\pi}\cdot{\bm\pi}\right)\nonumber \\
   &+&\frac{1}{2G'}\left({\bm V}_a\cdot{\bm V}_a+{\bm A}_a\cdot{\bm A}_a\right)
\end{eqnarray}
being $G'=G/2$ and
\begin{equation}
i {\cal D} = i {\slashed \partial} + {\hat e} \, {\slashed {\cal A}} - \hat m + {\slashed {\hat V}} + \gamma^5 {\slashed {\hat A}} - (\sigma +  i\gamma^5 {\bm\pi}\cdot{\bm\tau})\,. \quad
\label{eq:icD}
\end{equation}
Now, we are in a position to compare our Lagrangian with that used by Chernodub in \cite{Chernodub:2011mc}. We note that our model is characterized by the constants $G=g^2/2\tilde\sigma$ and $G'=G/2$. It is easy to evaluate $G$ using results given in \cite{Frasca:2012eq} that grant an excellent agreement between our Nambu-Jona-Lasinio model and the experimental data. We set $g=1.52$ and $\tilde\sigma=(0.44\ GeV)^2$ and so $G^{-1}\approx 0.17\ GeV^2$ and $G^{'-1}\approx 0.34\ GeV^2$. So, the critical field as estimated in \cite{Chernodub:2011mc} is
\begin{equation}
    eB_c=\frac{9\pi^2}{2N_c}\left(\frac{1}{G'}-\frac{8}{9G}\right)
\end{equation}
and this gives the estimation $eB_c\approx 2.8\ GeV^2$, very near the approximate figure given in \cite{Chernodub:2011mc}. We have neglected the contribution coming from renormalization as we assume it to be at least one order of magnitude smaller than $1/G'$ mostly due to numerical factors.

\section{Magnetic field and temperature}
\label{sec4}

Our aim is to follow a different track to show the existence of a quantum phase transition in QCD. The idea is to generalize known results for the extended Nambu-Jona-Lasinio model (the one with vector interactions) to the case of finite temperature and magnetic field. The $\rho$ mass is seen to develop a singular behavior increasing the magnetic field.

One can evaluate the one-loop correction to the potential of the model given in eq.(\ref{eq:LB}) writing \cite{Bernard:1993rz}
\begin{eqnarray}
    V(\sigma,{\bm\pi},{\bm V}_a,{\bm A}_a)&=&-i{\rm Trln}\left(1+(i {\slashed \partial} + {\hat e} \, {\slashed {\cal A}} - \hat m)^{-1}
    ({\slashed {\hat V}} + \gamma^5 {\slashed {\hat A}} - (\sigma +  i\gamma^5 {\bm\pi}\cdot{\bm\tau}))\right) \nonumber \\
    &+&\int d^4x\left[\frac{1}{2}(G^{-1}+m_0^2)\sigma^2+\frac{1}{2G}{\bm\pi}\cdot{\bm\pi}\right]
    -\frac{1}{2G'}\int d^4x\left({\bm V}_a\cdot{\bm V}_a+{\bm A}_a\cdot{\bm A}_a\right)
\end{eqnarray}
being in our case \cite{Ritus:1972ky,Leung:2005yq}
\begin{equation}
\label{eq:QP}
   {\rm Tr}(i {\slashed \partial} + e_f \, {\slashed {\cal A}} - m_f)^{-1}=
   S_f(x,y)= \sum_{k=0}^\infty\int\frac{dp_0 dp_2 dp_3}{(2\pi)^4}
   E_P(x)\Lambda_k \frac{i}{{\slashed P} - m_f}\bar{E}_P(y)~,
\end{equation}
where $f=(u,d)$, $E_P(x)$ corresponds to the eigenfunction of a charged fermion in magnetic field, and $\bar{E}_P(x) \equiv
\gamma_0(E_P(x))^\dagger \gamma_0$. In the above equation,
\begin{equation}
P = (p_0,0,{\cal Q}\sqrt{2k|e_fB|},p_3)~,\label{eq:MB}
\end{equation}
where $k =0,1,2,\dots$ labels the $k^{\text{th}}$ Landau level, and ${\cal Q} \equiv\text{sign}(e_f)$, with $e_f$ denoting the
charge of the flavor $f$ and $m_f$ the corresponding mass; $\Lambda_k$ is a projector in Dirac space which keeps into account the degeneracy of the Landau levels; it is given by
\begin{equation}
\Lambda_k = \delta_{k0}\left[{\cal P_+}\delta_{{\cal Q},+1} +
{\cal P_-}\delta_{{\cal Q},-1}\right] + (1-\delta_{k0})I~,
\end{equation}
where ${\cal P}_{\pm}$ are spin projectors and $I$ is the identity matrix in Dirac spinor indexes. The one-loop effective potential can be obtained by observing that we have a sum on the Landau levels (accounting for their degeneracy). Moving to a temperature dependence goes as usual \cite{Das:1997gg}. 
We follow a somewhat different approach than that seen in the current literature observing that, in our case, a dimensional reduction occurs
\begin{equation}
   \int\frac{d^4p}{(2\pi)^4}\rightarrow\frac{eB}{2\pi}\sum_k\beta_k\int\frac{dp_0dp_3}{2\pi}.
\end{equation}
being $\beta_k=2-\delta_{k0}$ the degeneracy of the Landau levels. So, let us consider the contribution due to the quarks and one has
\begin{eqnarray}
   V_q&=&-i{\rm Tr}\ln\left(i\slashed\partial-e\slashed A +\sigma+i\gamma^5\tau\pi
   +\gamma^\mu\tau\cdot{\bm V}_\mu+\gamma_5\gamma^\mu\tau\cdot{\bm A}_\mu\right) \nonumber \\
   &-&(G^{-1}+m_0^2)\frac{\sigma^2}{2}-\frac{{\bm\pi}_a\cdot{\bm\pi}_a}{2G}+\frac{{\bm V}_a\cdot{\bm V}_a+{\bm A}_a\cdot{\bm A}_a}{2G'}.
\end{eqnarray}
This means that the contribution due to the one-loop at the effective potential becomes
\begin{equation}
   V_{1L}=-i{\rm Tr}\ln\left(i\slashed\partial-e\slashed A + v\right)=-\sum_f\frac{|e_fB|}{2\pi}\sum_k\beta_k\int\frac{dp_0dp_3}{(2\pi)^2}
   \ln(p_0^2+p_3^2+2k|e_fB|+v^2)
\end{equation}
where we have assumed a non-null vacuum expectation value for the $\sigma$ field, $v$. From this, one gets the corrections to the thermodynamic potential by summing over $p_0$ changed into a Matsubara sum. This will yield the well-known result \cite{Menezes:2008qt,Fukushima:2012vr,Allen:2013lda}
\begin{eqnarray}
\Omega_f(\mu,B,T) &=& (G^{-1}+m_0^2)\frac {\sigma^2}{2}
 - N_c\sum_f\frac{|e_fB|}{2\pi}\sum_k\beta_k\int \frac{d p_z}{(2\pi)}   \omega_k(p_z)  \nonumber \\
&-&N_cT\sum_f\frac{|e_fB|}{2\pi}\sum_k\beta_k\int \frac{d p_z}{(2\pi)} \left \{\ln[ 1+ e^{-[\omega_k(p_z)+\mu_f]/T}] \right . \nonumber \\
&+& \left .\ln [1+ e^{-[\omega_k(p_z)-\mu_f]/T}] \right \} \,\,\,,
\label{BTmu}
\end{eqnarray}
where $\omega_k(p_z)=\sqrt{p_z^2+2k|e_fB|+v^2}$ and $\mu_f$ is the chemical potential. The vacuum contribution is clearly diverging and need a regularization. From this potential a gap equation can be derived and it can be shown that the critical temperature is seen to increase with increasing magnetic field.

Our aim is to compute the correction to the potential arising from the vector part of the Nambu-Jona-Lasinio model. This computation is widely known since the nineties \cite{Ebert:1994mf} and give for the $\rho$ mass
\begin{equation}
    m_\rho^2=\frac{3}{8G'N_cJ_2(0)}
\end{equation} 
being
\begin{equation}
    J_2(0)=-iN_f\int\frac{d^4p}{(2\pi)^4}\frac{1}{(p^2-v^2+i\epsilon)^2}
\end{equation}
that is clearly divergent. This can be regularized in different ways and its value is $(1/8\pi^2)\ln(\Lambda^2/v^2)$. A straightforward computation gives an estimation for the $\rho$ mas of about $670\ MeV$ but this can be significantly improved adding further corrections \cite{Klevansky:1997dk}. Now, we move to thermodynamics and apply a magnetic field so that the integration rules are now \cite{Menezes:2008qt}
\begin{eqnarray}
    \int\frac{d^4p_E}{(2\pi)^4}&\rightarrow& T\sum_f\frac{|e_fB|}{2\pi}\sum_k\beta_k\sum^{\infty}_{n=-\infty}\int\frac{dp_z}{2\pi} \nonumber \\
    p_z^2&\rightarrow&p_z^2+2k|e_fB| 
\end{eqnarray}
and we will get
\begin{equation}
    J_2(B,T)=T\sum_f\frac{|e_fB|}{2\pi}\sum_k\beta_k\sum^{\infty}_{n=-\infty}\int\frac{dp_z}{2\pi}\frac{1}{[(2n+1)^2\pi^2T^2+p_z^2+2k|e_fB|+v^2]^2}
\end{equation}
where now $v=v(B,T)$ is the effective quark mass. Matsubara sum can be immediately performed giving
\begin{equation}
   \sum^{\infty}_{n=-\infty}\frac{1}{[(2n+1)^2\pi^2T^2+p_z^2+2k|e_fB|+v^2]^2}=\frac{1}{4T}\frac{\cosh\left(\frac{\omega_k(p_z)}{2T}\right)
   \sinh\left(\frac{\omega_k(p_z)}{2T}\right)-\frac{\omega_k(p_z)}{2T}}{\omega^{3}_k(p_z)\cosh^2\left(\frac{\omega_k(p_z)}{2T}\right)}.
\end{equation}
For  small $T$ this sum reduces to
\begin{equation}
   \sum^{\infty}_{n=-\infty}\frac{1}{[(2n+1)^2\pi^2T^2+p_z^2+2k|e_fB|+v^2]^2}\approx
   \frac{1}{4T}\frac{1-\frac{2\omega_k(p_z)}{T}e^{-\left(\frac{\omega_k(p_z)}{T}\right)}}{\omega^{3}_k(p_z)}.
\end{equation}
This yields upon integration on $p_z$
\begin{equation}
   J_2(B,T)\approx \sum_f\frac{|e_fB|}{16\pi^2}\sum_k\beta_k\left[\frac{1}{2k|e_fB|+v^2}-\frac{2}{T}
   \frac{F\left(\frac{\sqrt{2k|e_fB|+v^2}}{T}\right)}{\sqrt{2k|e_fB|+v^2}}\right]
\end{equation}
and
\begin{equation}
   F(a) = \pi-2a\left[K_0(a)+\frac{\pi}{2}(K_0(a){\bm L}_1(a)+K_1(a){\bm L}_0(a))\right]
\end{equation}
where $K_n$ are Macdonald functions and ${\bm L}_n$ modified Struve functions. We note that $\lim_{a\rightarrow\infty}F(a)=0$ exponentially. The last step is to regularize the harmonic series as
\begin{equation}
\label{eq:exact}
    \sum_{k=0}^{\infty}\beta_k\frac{1}{2k|e_fB|+v^2}=-\frac{1}{v^2}+\frac{1}{|e_fB|}S(B,v)
		=-\frac{1}{v^2}-\frac{1}{|e_fB|}\psi\left(\frac{v^2}{2|e_fB|}\right)
\end{equation}
where $S(B,v)=\sum_{k=0}^{\infty}\frac{1}{k+\frac{v^2}{2|e_fB|}}$ and $\psi(x)$ is the digamma function. But in order to recover the limit of magnetic field going to zero, we have to add a cut-off to this series as is customary \cite{Ruggieri:2013cya}. We put
\begin{eqnarray} 
\frac{1}{|e_fB|}S(B,v)&=&-\frac{1}{v^2}+\frac{1}{|e_fB|}\psi\left(\frac{v^2}{2|e_fB|}+1+N\right)
-\frac{1}{|e_fB|}\psi\left(\frac{v^2}{2|e_fB|}\right)= \nonumber \\
&-&\frac{1}{v^2}+\frac{1}{|e_fB|}\ln N
-\frac{1}{|e_fB|}\psi\left(\frac{v^2}{2|e_fB|}\right) \nonumber \\
&=&\frac{1}{|e_fB|}\ln\left(\frac{\Lambda^2}{v^2}\right)+\frac{1}{3}\frac{|e_fB|}{v^4}+\ldots
\end{eqnarray}
as our regularization procedure implies that $2N|e_fB|=\Lambda^2$, the physical cut-off. This gives back the right mass for $\rho$ as we recover the proper value $J_2(0)=(1/8\pi^2)\ln(\Lambda^2/v^2)$ as it should. But we are interested in the opposite limit for a strong magnetic field and we use a different expansion. We will get
\begin{equation}
   \frac{1}{|e_fB|}S(B,v)\approx\frac{1}{v^2}+\frac{\gamma}{|e_fB|}-\frac{\pi^2}{6}\frac{v^2}{2|e_fB|^2}
\end{equation}
So, finally we can collect all these computations into the equation
\begin{eqnarray}
   J_2(B,T)&\approx&\frac{1}{16\pi^2}\sum_f\left[\frac{|e_fB|}{v^2}+\gamma-\frac{\pi^2}{6}\frac{v^2}{2|e_fB|}\right. \nonumber \\
   &-&\left.\frac{2|e_fB|}{T}
   \sum_{k=0}^\infty\frac{F\left(\frac{\sqrt{2k|e_fB|+v^2}}{T}\right)}{\sqrt{2k|e_fB|+v^2}}\right].
\end{eqnarray}
Already in the asymptotic approximation, we can check the existence of a critical field at $T=0$ for $\rho$ mass to change sign and the validity of the vector meson dominance as $m_\rho\rightarrow 0$ as $B\rightarrow\infty$. This is due to the fact the exact sum (\ref{eq:exact}) has indeed a positive zero on the real axis for the ratio $|eB|/v^2$ at about $0.9787896\ldots$. This proves the existence of a quantum pahse transition in QCD at increasing magnetic field and lowering temperature.

\section{Conclusions}
\label{sec5}

We have shown that a quantum phase transition occurs in the ground state of QCD when a strong magnetic field is applied. A condensate of $\rho$ mesons can form in agreement with a recent proposal. We were able to get the full formula for the $\rho$ mass with a magnetic field and with temperature going to zero but not null. As a by-product we were able to get a closed formula for the critical magnetic field at which the transition occurs. This transition is purely quantum as it is obtained when the magnetic field is varied rather then the temperature. The importance of this result stems from the fact that we showed how a Nambu-Jona-Lasinio model is straightforwardly obtained starting from the QCD Lagrangian. This is possible because we were able to get a closed formula for the gluon propagator in the infrared limit. We hope this should pave the way for a possible experimental test of this fascinating possibility.






\end{document}